\documentclass{article}
\usepackage{amssymb,amsmath,amsxtra}
\usepackage{graphics}
\usepackage{epsfig}
\usepackage{axodraw}

\usepackage{float}
\allowdisplaybreaks[4]

\newcommand{\diff}{{\rm{d}}}
\newcommand{\ma}{{\mathcal{A}}}
\newcommand{\ml}{{\mathcal{L}}} 

\begin{document}
\title{Wrong-Helicity Electrons in Radiative Muon Decay}

\author{V.~S.~Schulz and L.~M.~Sehgal}
\date{\it Institute of Theoretical Physics, RWTH Aachen \\ D-52056 Aachen, Germany}

\maketitle

\begin{abstract}

We have studied in detail the spectrum of the decay
$\mu^- \to e^- \bar{\nu}_e \nu_\mu \gamma$ as a function of the 
electron helicity, and verify the prediction \cite{Sehgal:2003mu} that
a significant fraction of the electrons in this decay is right-handed. 
These ``wrong-helicity'' electrons persist in the limit
$\lambda = m_e / m_\mu \to 0$, and are connected with helicity-flip
bremsstrahlung in QED. The longitudinal polarization of the electron
is calculated as a function of the photon and electron energy, and
deviates systematically from the naive V-A prediction $P_L = -1$. 
The right-handed component is concentrated in the collinear region 
$\theta \lesssim m_e / E_e$. In the limit $\lambda \to 0$, 
we reproduce the results obtained in \cite{Sehgal:2003mu} using the
helicity-flip splitting function introduced by Falk and Sehgal.
\end{abstract}

\section{Introduction}

In a recent Letter \cite{Sehgal:2003mu} it was pointed out that as a consequence of
helicity-flip bremsstrahlung, electrons in the radiative decay 
$\mu^- \to e^- \bar{\nu}_e \nu_\mu \gamma$ are not purely left-handed,
even in the limit $m_e \to 0$. The decay width into right-handed
electrons was shown to be $\Gamma_R = \frac{\alpha}{4 \pi} \Gamma_0$, 
where $\Gamma_0 \equiv G_{\rm{F}}^2 m_\mu^5 / (192 \pi^3)$.
The energy spectrum of these wrong-helicity electrons was calculated,
as also that of the photons accompanying them. These results were
obtained using the helicity-flip splitting function 
$D_{\rm{hf}} (z) = \frac{\alpha}{2 \pi} z$ introduced by Falk and Sehgal
\cite{Falk:1993tf}.

The appearance of right-handed electrons in $\mu$-decay is, at first
sight, surprising since it goes against the conventional wisdom based
on the V-A structure of weak interaction and the common assumption of
helicity-conservation in massless QED. However persistence of
helicity-flip bremsstrahlung in the $m_e \to 0$ limit has been established 
in several calculations of radiative processes 
\cite{Contopanagos:ga,Jadach:1987ws,Trentadue_Smilga,Groote:1995yc,Fischer:zh}, 
going back to the seminal paper of Lee and Nauenberg \cite{Lee:is}. 
In all cases, the origin of the effect is the characteristic behaviour
of helicity-flip bremsstrahlung at small angles
\begin{equation}
\label{cross-section-1}
    \diff \sigma_{\rm{hf}} \approx 
    \Big( \frac{m_e}{E_e} \Big)^2 
    \frac{\diff \theta^2}
    {\big[ \theta^2+ \big( \frac{m_e}{E_e} \big)^2 \big]^2} \,,
\end{equation}
leading to a non-zero integrated rate that survives the limit $m_e \to0$.

In the present paper, we derive the helicity-dependent spectrum of the decay
$\mu^- \to e^- \bar{\nu}_e \nu_\mu \gamma$ from first principles,
without neglecting the electron mass. We obtain the longitudinal
polarization $P_{\it{L}}$ of the electron as a function of the electron energy,
photon energy and the angle between the electron and photon. We then
consider the limit $m_e \to 0$ to demonstrate that the deviation from 
$P_{\it{L}} = -1$ is present even in the chiral limit of QED.
The results are compared with those obtained in the equivalent
particle approach based on the helicity-flip fragmentation function, 
thus providing a confirmation of the statements in Ref.\cite{Sehgal:2003mu}.

\section{Matrix Element and Phase Space for $\mu^- \to e^- \bar{\nu}_e \nu_\mu \gamma$}

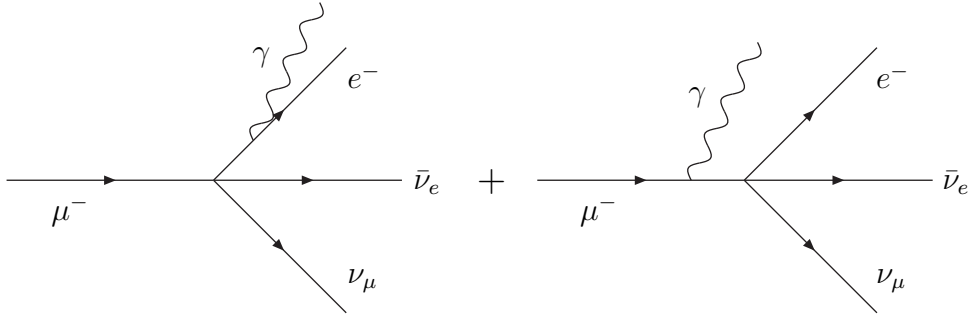
\begin{figure}[ht]
\begin{center}
\begin{picture}(370,120)
\SetWidth{0.5}
\SetColor{Black}
\ArrowLine(78,50)(149,50)
\ArrowLine(78,50)(128,0)
\ArrowLine(0,50)(78,50)
\ArrowLine(78,50)(128,100)
\Photon(93,65)(118,117){3}{4}
\Text(17,33)[lb]{\large{$\mu^-$}}
\Text(93,93)[lb]{\large{$\gamma$}}
\Text(129,85)[lb]{\large{$e^-$}}
\Text(154,45)[lb]{\large{${\bar{\nu}}_e$}}
\Text(129,8)[lb]{\large{$\nu_{\mu}$}}
\Text(178,45)[lb]{\Large{+}}
\ArrowLine(278,50)(349,50)
\ArrowLine(278,50)(328,0)
\ArrowLine(200,50)(278,50)
\ArrowLine(278,50)(328,100)
\Photon(258,50)(283,102){3}{4}
\Text(217,33)[lb]{\large{$\mu^-$}}
\Text(258,78)[lb]{\large{$\gamma$}}
\Text(329,85)[lb]{\large{$e^-$}}
\Text(354,45)[lb]{\large{${\bar{\nu}}_e$}}
\Text(329,8)[lb]{\large{$\nu_{\mu}$}}
\end{picture}
\caption{Feynman diagrams \label{FeynDia}}
\end{center}
\end{figure}

The matrix element for the decay 
$\mu^- \to e^- \bar{\nu}_e \nu_\mu \gamma$ may be obtained from the
Feynman diagrams shown in Fig.(\ref{FeynDia}).
Adopting the notation of Ref.\cite{Fischer:2002hn,Fischer:pn}, we obtain
the square of the matrix element, summed over the photon polarization, taking
the muon to be unpolarized.
The spectrum is derived in terms of the invariants
\begin{equation}
    x \,:=\, \frac{2\,p_\mu\!\cdot\!p_e}{m_\mu^2} \,, \qquad
    y \,:=\, \frac{2\,p_\mu\!\cdot\!k}{m_\mu^2} \,, \qquad
    z \,:=\, \frac{2\,p_e\!\cdot\!k}{m_\mu^2}
\end{equation}
and the mass ratio $\lambda := m_e / m_\mu$. In the rest frame of the muon,
\begin{equation}
    x \,=\, \frac{2 E_e}{m_\mu} \,, \quad 
    y \,=\, \frac{2 E_\gamma}{m_\mu} \,, \quad 
    z \,=\, \frac{y}{2}\, 
    [x - \cos(\theta)\,\sqrt{x^2 - 4 \lambda^2}] \,,
\end{equation}
$E_e, E_\gamma$ being the energies of the electron, photon and
$\cos (\theta)$ the angle between them. 
The boundary of phase space is determined by the condition
$-1 \le \cos (\theta) \le 1$, and 
$0 \le Q^2 \le (1 - \lambda)^2\,m_\mu^2$,
$Q^2$ being the ``missing mass'' of the
neutrino pair. This leads to the condition
\begin{equation}
    0 \le \, y \le y_{\rm{m}} (x) \,, \qquad
    2\,\lambda \le \, x \le 1 + \lambda^2 \,, \\
\end{equation}
where
\begin{equation}
\label{y_m-definition}
    y_{\rm{m}} (x) \,=\, 2\, 
    \frac{1 + \lambda^2 - x}{2 - x + \sqrt{x^2 - 4\,\lambda^2}} \,.
\end{equation}
Equivalently, the limits of phase space may be expressed as
\begin{equation}
    2\,\lambda \le \,x \le x_{\rm{m}} (y) \,, \qquad
    0 \le \,y \le 1 - \lambda \,, \\
\end{equation}
where
\begin{equation}
    \label{x_m-definition}
    x_{\rm{m}} (y) \,=\, \frac{\lambda^2 + (1 - y)^2}{1 - y} \,.
\end{equation}
The differential decay width for
$\mu^- \to e^- \bar{\nu}_e \nu_\mu \gamma$
in the variables $x$, $y$ and $\cos (\theta)$ is then given by
\begin{equation}
\label{dgamma1}
    \Big( 
    \frac{\diff \Gamma}{\diff x\,\diff y\,\diff \cos (\theta)}
    \Big)_{\rm{R,L}}
    = (-1)\,\Gamma_{0}\,\frac{\alpha}{2\,\pi}\,
    \sqrt{x^2 - 4\,\lambda^2}\,\frac{1}{y\,z^2}\,
    \frac{1}{2}\,(g_0 \pm g_1) \,
\end{equation}
where R and L denote right-handed and left-handed electrons,
corresponding to the polarization vectors
\begin{equation}
    {(s_e)}_{\rm{R}} =
    	\Big(
   \frac{|\mathbf{p_e}|}{m_e},\,\frac{E_e}{m_e}\,\mathbf{\hat{p}_e} 
    	\Big) \, \qquad
    {(s_e)}_{\rm{L}} =
    \Big(
    \frac{|\mathbf{p_e}|}{m_e},\,-\frac{E_e}{m_e}\,\mathbf{\hat{p}_e} 
    \Big) \,.
\end{equation}
The functions $g_0$ and $g_1$ are given by
\begin{equation}
\begin{split}
    g_0 := &-2\,x\,y\,z^2 + 6\,x\,y\,z^3 - 6\,x\,y^2\,z - 8\,x\,y^2\,z^2 + 6\,x\,y^3\,z 
    + 6\,x\,z^2 +\,8\,x\,z^3 - 6\,x^2\,y\,z - 8\,x^2\,y\,z^2 \\
    &+ 8\,x^2\,y^2\,z - 4\,x^2\,z^2 + 4\,x^3\,y\,z +\,6\,y\,z^2 + 5\,y\,z^3 
    - 2\,y\,z^4 - 2\,y^2\,z^2 + 2\,y^2\,z^3 - 3\,y^3\,z -\,2\,y^3\,z^2 \\
    &+ 2\,y^4\,z - 6\,z^3 - 4\,z^4 \\
    &+\lambda^2\,(8\,x\,y\,z + 6\,x\,y\,z^2 + 2\,x\,y^2\,z + 6\,x\,y^2 -\,8\,x\,y^3 
    + 6\,x\,z^2 - 6\,x^2\,y\,z - 4\,x^2\,y^2 + 6\,y\,z^2 - 3\,y\,z^3 \\ 
    &\phantom{+\lambda^2\,(}-6\,y^2\,z - 2\,y^2\,z^2 + 5\,y^3\,z + 6\,y^3 - 4\,y^4 - 8\,z^2 - 6\,z^3) \\
    & +\lambda^4\,(6\,x\,y^2 - 6\,y^2\,z - 8\,y^2 + 6\,y^3) \,, 
\end{split}
\end{equation}
\begin{equation}
\begin{split}
\label{def-g1}
    g_1 := 
    \big\{
    &6\,x^3\,y\,z - 4\,x^4\,y\,z + 6\,x^2\,y^2\,z - 8\,x^3\,y^2\,z + 3\,x\,y^3\,z - 6\,x^2\,y^3\,z 
    - 2\,x\,y^4\,z - 6\,x^2\,z^2 + 4\,x^3\,z^2 \\
    &- 6\,x\,y\,z^2 + 2\,x^2\,y\,z^2 + 8\,x^3\,y\,z^2 + 2\,x\,y^2\,z^2 + 8\,x^2\,y^2\,z^2
    + 2\,x\,y^3\,z^2 + 6\,x\,z^3 - 8\,x^2\,z^3 \\
    &- 5\,x\,y\,z^3 - 6\,x^2\,y\,z^3 - 2\,x\,y^2\,z^3 + 4\,x\,z^4 + 2\,x\,y\,z^4  \\
    &+ \lambda^2\,(-6\,x^2\,y^2 + 4\,x^3\,y^2 - 6\,x\,y^3 + 8\,x^2\,y^3 - 6\,y^4 +
    8\,x\,y^4 - 24\,x\,y\,z + 16\,x^2\,y\,z - 4\,y^2\,z^3 \\
    &\phantom{+ \lambda^2\,(} + 4\,y^5 + 2\,x^3\,y\,z + 22\,x\,y^2\,z - 6\,x^2\,y^2\,z + 4\,y^3\,z 
    - 7\,x\,y^3\,z - 4\,y^4\,z + 24\,z^2 - 16\,x\,z^2 - 2\,x^2\,z^2  \\
    &\phantom{+ \lambda^2\,(}- 16\,y\,z^2 - 18\,x\,y\,z^2 - 2\,x^2\,y\,z^2 - 6\,y^2\,z^2 +
    10\,x\,y^2\,z^2 + 4\,y^3\,z^2 + 16\,z^3 + 2\,x\,z^3 + x\,y\,z^3 )  \\
    &+ \lambda^4\,(24\,y^2 - 16\,x\,y^2 - 2\,x^2\,y^2 - 16\,y^3 - 2\,x\,y^3 - 2\,y^4
    - 8\,x\,y\,z + 16\,y^2\,z + 2\,x\,y^2\,z - 4\,y^3\,z \\
    &\phantom{+ \lambda^4\,(}+ 8\,z^2 - 2\,y^2\,z^2) \\
    &+ 8\,\lambda^6\,y^2
    \big\} \cdot
    \frac{1}{2\,{\sqrt{x^2 - 4\,{\lambda}^2}}} \,.
\end{split}
\end{equation}
Adding the expressions for $\diff \Gamma_{\rm{R}}$ and $\diff \Gamma_{\rm{L}}$, 
we obtain the spectrum for unpolarized electrons, 
which depends on the function $g_0$  only \cite{Fischer:pn}.
The specific effects of electron helicity are contained in the
function $g_1$.

\section{Unpolarized Spectra (summed over electron helicity)}\label{UnpolSpectra}

Summing over the electron helicity, the differential decay rate of
$\mu^- \to e^- \bar{\nu}_e \nu_\mu \gamma$ is
\begin{equation}
    \Big( 
    \frac{\diff \Gamma}{\diff x\,\diff y\,\diff \cos(\theta)}
    \Big)_{\rm{R+L}}
    = (-1)\,\Gamma_{0}\,\frac{\alpha}{2\,\pi}\,
    \sqrt{x^2 - 4\,\lambda^2}\,\frac{1}{y\,z^2}\,g_0 \,.
\end{equation}
From this we can derive the following spectra.

\subsection{Dalitz plot in electron and photon energies}

\begin{equation}
\begin{split}
\label{Dalitz-RL-1}
    \Big( 
    \frac{\diff \Gamma}{\diff x\,\diff y} 
    \Big)_{\rm{R+L}} &= 
    (-1)\,\Gamma_{0}\,\frac{\alpha}{2\,\pi} \,
    \int_{-1}^{+1}\,\diff \cos(\theta)\,
    \sqrt{x^2 - 4\,\lambda^2}\,
    \frac{1}{y\,z^2}\,g_0(x,y,z;\lambda) \\
    &= 
    \Gamma_{0}\,\frac{\alpha}{6\,\pi}\,\frac{1}{y}\,
    \Big(
    \ma\,
    \big\{ 
    12\,y\,(-6 + 3\,y + y^2) + x\,
    (-72 + 78\,y + 33\,y^2 - 6\,y^3) \\
    &\phantom{=\Gamma_{0}\,\frac{\alpha}{6\,\pi}\,\frac{1}{y}\,\Big(\ma\,\big\{}
    + 2\,x^2\,(24 + 12\,y - 5\,y^2 + 2\,y^3) \\
    &\phantom{=\Gamma_{0}\,\frac{\alpha}{6\,\pi}\,\frac{1}{y}\,\Big(\ma\,\big\{}
    + {\lambda}^2\,[96 - 72\,y + 4\,y^2 - 4\,y^3 + 9\,x\,(-8 - 2\,y + y^2)]  
    \big\} \\
    &\phantom{=\Gamma_{0}\,\frac{\alpha}{6\,\pi}\,\frac{1}{y}\,\Big(}+ 12\,\ml\,
    \big\{
    -4\,x^3 + x^2\,(6 - 8\,y) + 6\,{\lambda}^4\,y - 6\,x\,(-1 + y)\,y + (3 - 2\,y)\,y^2 \\
    &\phantom{=\Gamma_{0}\,\frac{\alpha}{6\,\pi}\,\frac{1}{y}\,\Big(+ 12\,\ml\,\big\{} 
    +{\lambda}^2\,[6\,x^2 + (6 - 5\,y)\,y - 2\,x\,(4 + y)]  
    \big\}
    \Big)\,,
\end{split}
\end{equation}
where we have defined
\begin{equation}
    \ma := \sqrt{x^2 - 4\,\lambda^2} \,,\quad
    \ml := \frac{1}{2}\,\log \Big( \frac{x + \ma}{x - \ma} \Big) \,.  
\end{equation}

\subsection{Photon Energy Spectrum}

The photon energy spectrum $(\diff \Gamma / \diff y)_{\rm{R+L}}$ is given 
explicitly in Eq.(\ref{PS-RL}) of the Appendix
The leading term for small $y$ (small photon energies) is
\begin{equation}
    \Big( \frac{\diff \Gamma}{\diff y} \Big)_{\rm{R+L}} 
    \xrightarrow{y \to 0}
    \Gamma_0\,\frac{\alpha}{6\,\pi}\,\frac{1}{y}\, 
    \big[
    (-17 + 64\,{\lambda}^2 - 64\,{\lambda}^6 + 17\,{\lambda}^8)
    + (-12 + 144\,\lambda^4 - 12\,\lambda^8)\,\log(\lambda) 
    \big] \,.
\end{equation}
In the limit of small $\lambda$, this reduces to the approximate form
\begin{equation}
    \Big(
    \frac{\diff \Gamma}{\diff y} 
    \Big)_{\rm{R+L}} 
    \xrightarrow{\lambda \to 0}
    \Gamma_0\,\frac{\alpha}{6\,\pi}\,\frac{1}{y}\, 
    \big[ 
    -17  - 12\, \log (\lambda) 
    \big] \,,
\end{equation}
which agrees with the old calculation of Eckstein and Pratt \cite{Eckstein} and
Kinoshita and Sirlin \cite{Kinoshita:1958ru}, in which terms of order $\lambda^2$ and
higher powers were neglected. The integrated rate for $y > y_0$ is
\begin{equation}
    \Gamma_{\rm{R+L}} (y_0;\lambda)  
    \xrightarrow{\lambda \to 0, \, y_0 \to 0}
    \Gamma_0\,\frac{\alpha}{\pi}\, 
    \Big[
    \log (\lambda) \Big( 2 \, \log (y_0) + \frac{7}{3} \Big)
    + \frac{17}{6}\,\log (y_0) - \frac{\pi^2}{3} + \frac{13409}{2520}
    \Big] \,, 
\end{equation}
which agrees with Ref. \cite{Eckstein}, except in the constant 
(non-logarithmic) term, which was \,$- \pi^2/6 + 601/144$\, in the old calculation.

\subsection{Electron energy spectrum}\label{ees}

The exact result for the electron energy spectrum 
$(\diff \Gamma / \diff x)_{\rm{R+L}}$ is given in Eq.(\ref{xi2-u-1}) of the Appendix.

\begin{figure}[h]
  \begin{center}
    \unitlength1truecm
\resizebox{8cm}{!}{
    \begin{picture}(9,5.5)
      \put(0,0){\psfig{file=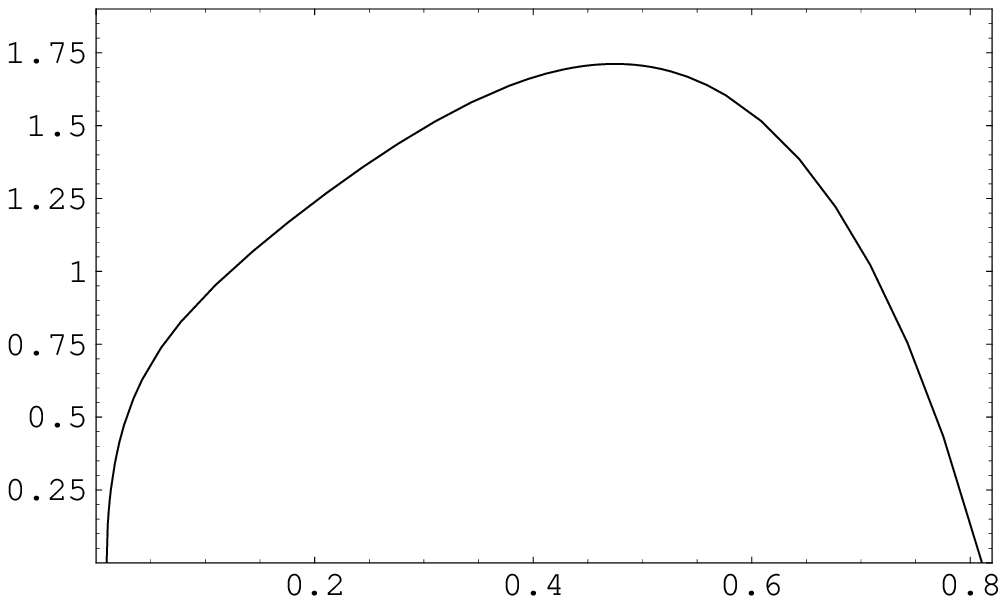,width=8.5cm}}
      \put(1.2,4.5){a}
      \put(4.5,-0.3){{x}}
    \end{picture}}
\resizebox{8cm}{!}{
    \begin{picture}(9,5.5)
      \put(0,0){\psfig{file=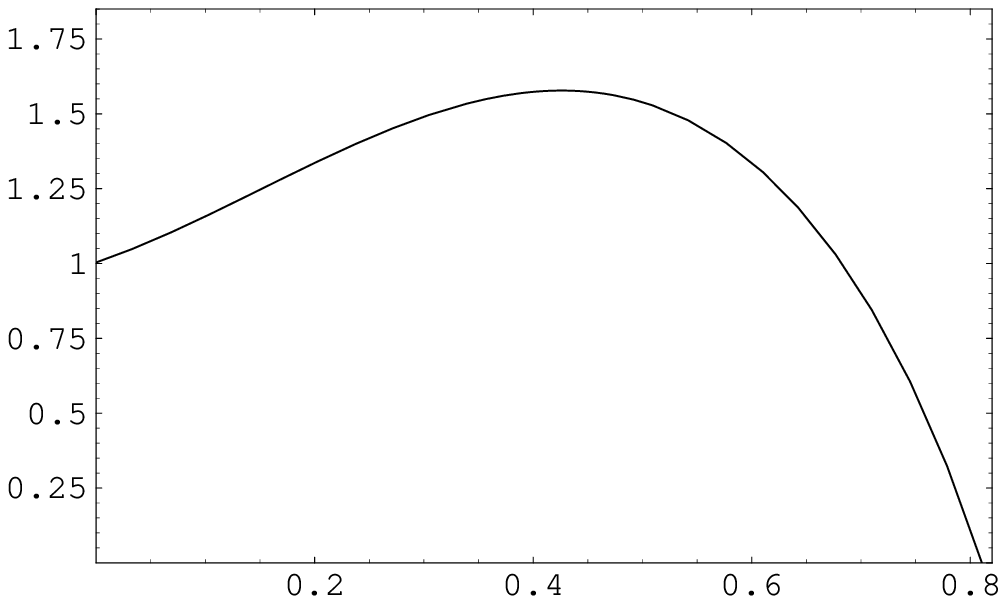,width=8.5cm}}
      \put(1.2,4.5){b}
      \put(4.5,-0.3){{x}}
    \end{picture}}
\end{center}
  \caption{{Normalized electron energy spectrum $(\Gamma_{\rm{R+L}})^{-1}\,
    \big(
    \frac{\diff \Gamma}{\diff x}
    \big)_{\rm{R+L}}$ for $y_0 \approx 0.189\,(E_{\gamma 0} =
    10\,\rm{MeV})$ and a) $\lambda = m_e/m_\mu \approx 1/207$, b) 
    $\lambda \to 0$.}
    \label{xspectrum1}} 
\end{figure}

The normalized electron energy spectrum is shown in Fig.(\ref{xspectrum1}a).
\noindent In the limit $\lambda \to 0$ we have for the $x$-spectrum
\begin{equation}
\begin{split}
\label{xs-n-limit-1}
    \lim_{\lambda \to 0}\,(\Gamma_{\rm{R+L}})^{-1}\,
    \Big(
    \frac{\diff \Gamma}{\diff x}
    \Big)_{\rm{R+L}} = 
    &\Big\{
    (1 - y_0 - x)\,[-5 - 5\,y_0 + 4\,{y_0}^2 + x\,(-17 + 14\,y_0) +
    34\,x^2] \\
    &\phantom{\Big\{}+ 12\,x^2\,(-3 + 2\,x)\,
    \log \Big( \frac{1 - x}{y_0} \Big)
    \Big\} \\
    &\times\,
    \Big[
    (1 - y_0)\,(7 - 3\,y_0 + 3\,{y_0}^2 - {y_0}^3) + 6\,\log (y_0)
    \Big]^{-1} \,.
\end{split}
\end{equation}
This is shown in Fig.(\ref{xspectrum1}b).

\noindent In the limit $y_0 \to 0$ we reproduce the spectrum for 
non-radiative muon decay 
\begin{equation}
\label{xs-n-limit-2}
    \lim_{\lambda \to 0,\,y0 \to 0}\,(\Gamma_{\rm{R+L}})^{-1}\,
    \Big(
    \frac{\diff \Gamma}{\diff x}
    \Big)_{\rm{R+L}} = 
    (\Gamma_{\rm{R+L}})^{-1}\,
    \Big( \frac{\diff \Gamma}{\diff x} 
    \Big)^{\rm non-rad} =
    2\,x^2\,(3 - 2\,x) \,. 
\end{equation}

\section{Spectra for Right- and Left-handed Electrons}

The spectra when the final electron is polarized in the state R or L
are obtained using Eq.(\ref{dgamma1}). We will emphasize the case R,
which is the one which behaves in an unexpected way in the $\lambda
\to 0$ limit \cite{Sehgal:2003mu,Fischer:2002hn}.
The corresponding spectra for the state L are
obtained by subtraction from the unpolarized result L+R 
discussed in Sec.\ref{UnpolSpectra}.

\subsection{Dalitz plot for R-Electrons}

\begin{equation}
\begin{split}
\label{dalitz-r1}
    \Big( \frac{\diff \Gamma}{\diff x\,\diff y} \Big)_{\rm{R}} = 
    \Gamma_0\,\frac{\alpha}{12\,\pi}\,\frac{1}{\ma}\,\frac{1}{y}\,
    &\big\{
    36\,\ma^2\,y^2  + \ma\,(36\,y^2 - 60\,x\,y^2 - 24\,y^3) + 
    12\,\ma\,{\lambda}^4\,(-8 + y^2) \\
    &\phantom{\big\{}+ {\lambda}^2\,
    \big[ 
    -288\,\ma + 96\,\ma^2 + 192\,\ma\,x - 72\,\ma^2\,x + 24\,\ma\,x^2 + 192\,\ma\,y  \\
    &\phantom{\big\{+ {\lambda}^2\,\big[} 
    - 72\,\ma^2\,y + 72\,\ma\,x\,y - 18\,\ma^2\,x\,y + 6\,\ma\,x^2\,y 
    + 48\,\ma\,y^2 + 4\,\ma^2\,y^2  \\
    &\phantom{\big\{+ {\lambda}^2\,\big[}  
    - 52\,\ma\,x\,y^2 + 9\,\ma^2\,x\,y^2 - 3\,\ma\,x^2\,y^2 -
    24\,\ma\,y^3 - 4\,\ma^2\,y^3 + 16\,\ma\,x\,y^3 
    \big]  \\
    &\phantom{\big\{}+ \ma\,(\ma - x)\,
    \big[ 
    -72\,x + 48\,x^2 - 72\,y + 78\,x\,y + 24\,x^2\,y + 33\,x\,y^2  \\
    &\phantom{\big\{+ \ma\,(\ma - x)\,\big[} 
    - 10\,x^2\,y^2 + 12\,y^3 - 6\,x\,y^3 + 4\,x^2\,y^3 
    \big]  \\  
    &\phantom{\big\{}+\ml\,
    \big[ 
    -12\,(\ma - x)\,(-3 + 2\,x + 2\,y)\,(2\,x^2 + 2\,x\,y + y^2)  \\
    &\phantom{\big\{+\ml\,\big[}
    + {\lambda}^4\,( 96\,x - 192\,y + 72\,\ma\,y - 24\,x\,y + 48\,y^2)   \\
    &\phantom{\big\{+\ml\,\big[} 
    + {\lambda}^2\,(288\,x - 96\,\ma\,x - 192\,x^2 + 72\,\ma\,x^2
    - 24\,x^3 + 72\,\ma\,y - 264\,x\,y  \\
    &\phantom{\big\{+\ml\,\big[+ {\lambda}^2\,(} 
    - 24\,\ma\,x\,y + 72\,x^2\,y - 48\,y^2 - 60\,\ma\,y^2 + 84\,x\,y^2 + 48\,y^3)  
    \big]
    \big\} \,. \\
\end{split}
\end{equation}
In the limit $\lambda \to 0$ we obtain
\begin{equation} 
\label{limit1}
    \lim_{\lambda \to 0}
    \Big( \frac{\diff \Gamma}{\diff x\,\diff y} \Big)_{\rm{R}} =  
    \Gamma_0\,\frac{\alpha}{\pi}\,y\,(3 - 2\,x - 2\,y) \,.
\end{equation}
This limiting result coincides with what one obtains on the basis of
the helicity-flip splitting function. Denoting the electron spectrum
in the non-radiative decay 
$(\diff \Gamma / \diff x_0)^{\rm{non-rad}}$,
the spectrum after helicity-flip bremsstrahlung is
\begin{equation}
    \Big( \frac{\diff \Gamma}{\diff x_0\,\diff z} \Big)_{\rm{R}} =  
    \Big( \frac{\diff \Gamma}{\diff x_0}
    \Big)^{\rm{non-rad}}\,D_{\rm{hf}} (z) \,.
\end{equation}
Noting that $y = x_0\,z,\,x = x_0\,(1 - z)$, and recalling that
\begin{equation}
    \Big(\frac{\diff \Gamma}{\diff x_0} \Big)^{\rm{non-rad}} =
    \Gamma_0\,2\,x_0^2\,(3 - 2\,x_0) \,, \quad  
    D_{\rm{hf}} (z) = \frac{\alpha}{2\,\pi}\,z \,, 
\end{equation}
we obtain
\begin{equation}
    \Big( \frac{\diff \Gamma}{\diff x\,\diff y} \Big)_{\rm{R}} =  
    \frac{1}{x_0}\,
    \Big( \frac{\diff \Gamma}{\diff x_0\,\diff z} \Big)_{\rm{R}} =
    \Gamma_0\,\frac{\alpha}{\pi}\,y\,(3 - 2\,x - 2\,y) \,,
\end{equation}
exactly as in Eq.(\ref{limit1}).

\subsection{Photon-Spectra for R- and L-electrons}\label{Photon-Spectra}

\begin{figure}[h]
  \begin{center}
    \unitlength1truecm
    \begin{picture}(8,4.96)
      \put(0,0){\psfig{file=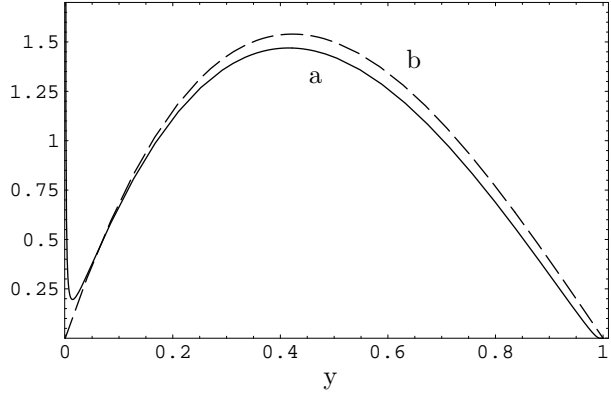,width=8cm}}
      \put(4,3.8){a}
      \put(5.3,4){b}
      \put(4.2,-0.2){y}
    \end{picture}
  \end{center}
  \caption{{Photon energy spectrum 
      $\big(
    \frac{\diff \Gamma}{\diff y} 
    \big)_{\rm{R}}$
      for right-handed electrons in units of $\Gamma_0\,\frac{\alpha}{4\,\pi}$ for
      a) $\lambda = 1/207$, b) $\lambda \to 0$.}
    \label{PSRplot1}} 
\end{figure}

\begin{figure}[h]
  \begin{center}
    \unitlength1truecm
    \begin{picture}(10,5)
      \put(0,0){\psfig{file=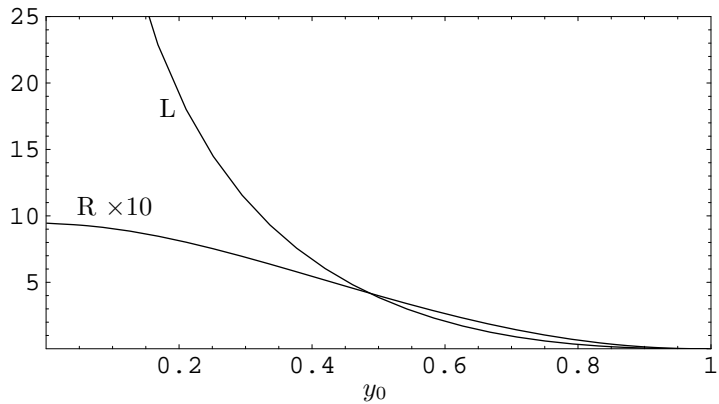,width=10cm}}
      \put(5,-0.2){{$y_0$}}
      \put(1.2,2.2){{R $\times 10$}}
      \put(2.3,3.5){L}
    \end{picture}
  \end{center}
  \caption{{Integrated rates for R- and L-electrons 
      $\Gamma_{\rm{R,L}} (y_0)$ in units of
      $\Gamma_0\,\frac{\alpha}{4\,\pi}$
      as function of minimum photon energy $y_0$.}
    \label{widthplot1}} 
\end{figure}

\begin{figure}[h]
  \begin{center}
    \unitlength1truecm
    \begin{picture}(10,6)
      \put(0,0){\psfig{file=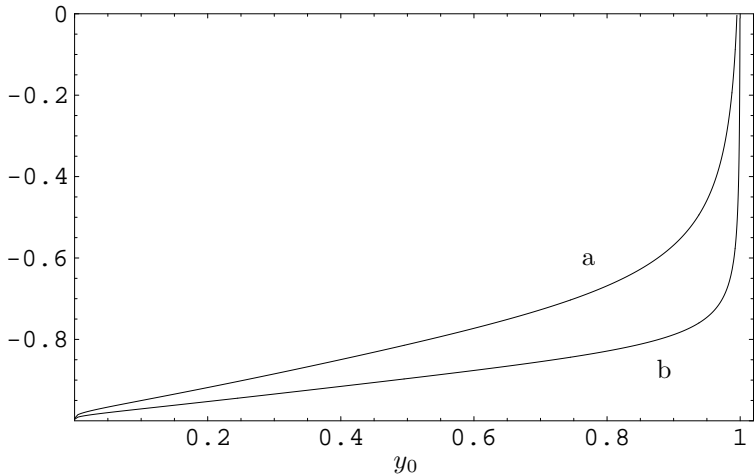,width=10cm}}
      \put(5.2,-0.2){{$y_0$}}
      \put(7.7,2.5){{{a}}}
      \put(8.7,1){{{b}}}
    \end{picture}
  \end{center}
  \caption{{Longitudinal polarization of electron $P_{\it{L}}$ as
      function
      of minimum photon energy $y_0$: a) $\mu \to e$ decay, b) $\tau
      \to e$ decay.}
    \label{polplot2}} 
\end{figure}

Of particular interest is the photon energy spectrum associated with
the right-handed polarization state of the electron, which is
naively forbidden in the $\lambda \to 0$ limit.
The corresponding spectrum for L-electrons may be obtained by
subtraction from Eq.(\ref{PS-RL}).
The exact result for 
$(\diff \Gamma / \diff y)_{\rm{R}}$ is stated in Eq.(\ref{photon-spectrum-R-eq}).
In the limit $m_e / m_\mu \to 0$ one obtains
\begin{equation}
\label{PSRlimit}
    \lim_{\lambda \to 0} \Big( \frac{\diff \Gamma}{\diff y} \Big)_{\rm{R}} = 
    \Gamma_0\,\frac{\alpha}{\pi}\,y\,(1 - y)\,(2 - y) \,,
\end{equation}
which is the result derived in \cite{Sehgal:2003mu} on the basis of the helicity-flip
fragmentation function $D_{\rm{hf}} (z)$. The spectrum for  $\lambda \neq 0$ is
plotted in Fig.(\ref{PSRplot1}).
\noindent The striking feature is the appearance of a divergence at $y = 0$.
This is connected with the fact that for $m_e \neq 0$, helicity is not
a good quantum number. As a consequence, ordinary (non-flip)
bremsstrahlung, diverging as $1/y$ at small $y$, is partly transmitted
to the final state with electron polarization $s_e = s_e^R$. This contribution
vanishes in the limit $\lambda \to 0$, leaving behind the anomalous
(helicity-flip) contribution in Eq.(\ref{PSRlimit}).
This anomalous bremsstrahlung integrates to a decay width 
\begin{equation}
\label{width-R-1}
    \Gamma_{\rm{R}} = \Gamma_0\,\frac{\alpha}{4\,\pi}\,
    (1 - y_0)\,(1 + y_0 - 3\,{y_0}^2 + {y_0}^3) \,,
\end{equation}
which is the contribution of right-handed electrons to the muon decay
width in the  $\lambda \to 0$ limit \cite{Sehgal:2003mu}.
The exact expression for $\Gamma_{\rm{R}} (y_0)$ and the corresponding left-handed rate
$\Gamma_{\rm{L}} (y_0)$ are plotted in Fig.(\ref{widthplot1}).
From these, we obtain the longitudinal polarization
\begin{equation}
    P_{\it{L}} (y_0) := \frac{\Gamma_{\rm{R}}(y_0) - \Gamma_{\rm{L}}(y_0)}
    {\Gamma_{\rm{R}}(y_0) + \Gamma_{\rm{L}}(y_0)}
\end{equation}
plotted in Fig.(\ref{polplot2}).
\noindent This figure demonstrates the presence of a significant right-handed electron
component in radiative muon decay, which becomes as probable as the
left-handed component at high photon energies.
To illustrate the approach to the chiral limit $\lambda \to 0$,
we have also shown in Fig.(\ref{polplot2}) the polarization of the
electron in the decay $\tau^- \to e^- \bar{\nu}_e \nu_\tau \gamma$.
In the chiral limit $\lambda \to 0$ the longitudinal polarization
goes from $P_{\it{L}} = -1$ to $P_{\it{L}} = 0$ like a step-function
at $y_0 = 1$.

\subsection{Electron Energy Spectra for R- and L-electrons}

The formula for the electron energy spectrum 
$(\diff \Gamma / \diff x)_{\rm{R}}$
for right-handed electrons is given in Eq.(\ref{PSR-1}), 
and may be compared with the unpolarized spectrum 
$(\diff \Gamma / \diff x)_{\rm{R+L}}$ in Eq.(\ref{xi2-u-1})
We have verified that in the limit $\lambda \to 0$, the right-handed
spectrum takes the form
\begin{equation}
    \Big(
    \frac{\diff \Gamma}{\diff x} 
    \Big)_{\rm{R}} \xrightarrow{\lambda \to 0}
    \Gamma_0 \, \frac{\alpha}{6\,\pi}\,(1 - y_0 - x)\,
    [5 + 5\,y_0 - 4\,{y_0}^2 + x\,(-7 - 2\,y_0) + 2\,x^2] \,.
\end{equation}
This is exactly the result derived in Ref.\cite{Sehgal:2003mu} on the
basis of the fragmentation function $D_{\rm{hf}} (z)$.

\section{Discussion of the Collinear Region}

\begin{figure}[h]
  \begin{center}
    \unitlength1truecm
      \begin{picture}(10,6.2)
        \put(0,0){\psfig{file=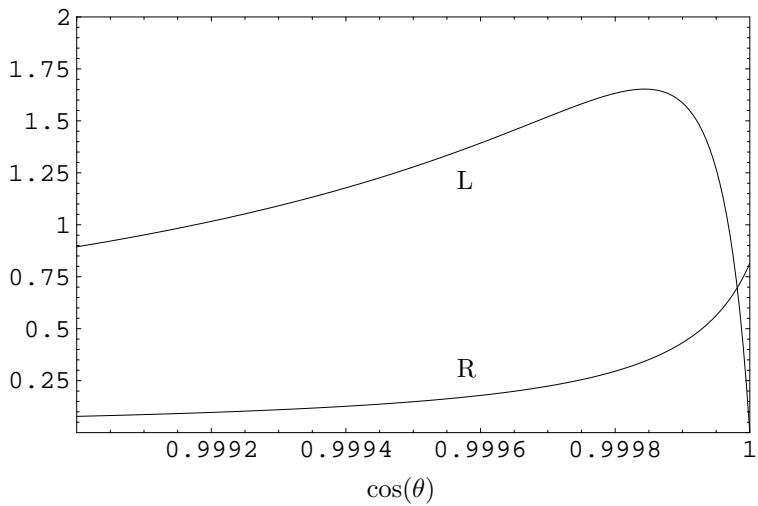,width=10cm}}
       \put(6,3.7){L}
       \put(6,1.2){R}
        \put(4.8,-0.4){$\cos (\theta)$}
      \end{picture}
  \end{center}
  \caption{{$\cos (\theta)$ spectrum 
      $\big(
      \frac{\diff \Gamma}{\diff \cos (\theta)} 
      \big)_{\rm{L,R}}$
      for $\lambda = 1/207$ and $y_0 = 0.189$ in units of $\Gamma_0$.}
    \label{CTplot1}} 
\end{figure}

\begin{figure}[h]
  \begin{center}
    \unitlength1truecm
      \begin{picture}(10,6.2)
        \put(0,0){\psfig{file=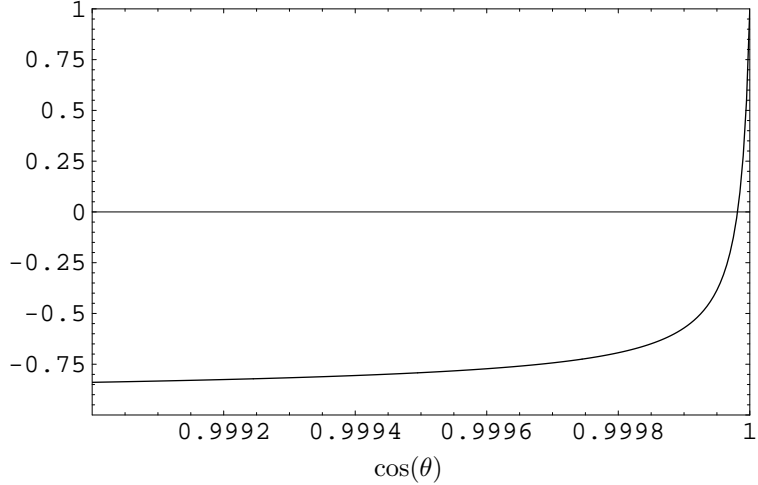,width=10cm}}
        \put(4.9,-0.3){$\cos (\theta)$}
      \end{picture}
  \end{center}
  \caption{{Longitudinal polarization $P_{\it{L}}$ 
      for $\lambda = 1/207$ and $y_0 = 0.189$.}
    \label{CTpolplot2}} 
\end{figure}

A question of interest is the behaviour of helicity-flip and
helicity-non-flip bremsstrahlung in the collinear region, 
$\theta \approx 0$, where $\theta$ is the angle between the photon and
electron in the final state. To this end, we have calculated exactly
the distribution $(\diff \Gamma / \diff \cos (\theta))_{\rm{R,L}}$.
The analytic results are, however, too lengthy to be written down here.
The left- and right-handed spectra for $\lambda = 1 / 207$, 
in the small angle region, are plotted in Fig.(\ref{CTplot1}).
\noindent The longitudinal polarization of the electron as a function of
$\theta$ is shown in Fig.(\ref{CTpolplot2}), and, changes from the V-A value
$P_{\it{L}} \approx -1$ to the value $P_{\it{L}} \approx +1$ in the forward
direction.
\noindent From the exact expressions we have derived the behaviour in the chiral
limit $\lambda \to 0$. For $\theta \ne 0$ and $\lambda \to 0$, we find
\begin{equation} 
\begin{split}
\label{CT-lim-R-1}
    \Big(
    \frac{\diff \Gamma}{\diff \cos (\theta)} 
    \Big)_{\rm{R}} \xrightarrow{\lambda \to 0}
    &\Gamma_0 \, \frac{\alpha}{6\,\pi}\,\pi\,\lambda
    \frac{1}{\theta^3}\,(1 - y_0)\,(5 + 5\,y_0 - 4\,{y_0}^2) \,, \\
    \Big(
    \frac{\diff \Gamma}{\diff \cos (\theta)} 
    \Big)_{\rm{L}} \xrightarrow{\lambda \to 0}
    &\Gamma_0 \, \frac{\alpha}{3\,\pi}\,
    \frac{1}{\theta^2}\,
    [-7 + 10\,y_0 - 6\,{y_0}^2 +
    4\,{y_0}^3 -{y_0}^4 - 6 \log (y_0)] \,.
\end{split}
\end{equation}
Note that, in the chiral limit, the $\lambda / \theta^3$ behaviour of
$(\diff \Gamma / \diff \cos (\theta))_{\rm{R}}$ refers to the opening angle of 
the electron and photon in the final state. The corrsponding $\lambda^2 / \theta^4$ behaviour of 
the helicity-flip scattering cross section (Eq.(\ref{cross-section-1})) refers to 
the angle between the photon and the incident electron.

We have also derived the value of 
$(\diff \Gamma / \diff \cos(\theta))$
in the forward direction ($\theta = 0$), for small $\lambda$:
\begin{equation} 
\begin{split}
\label{coll-lim-R-1}
    \Big( 
    \frac{\diff \Gamma}{\diff \cos(\theta)}
    \Big)_{\rm{R}} \bigg|_{\theta = 0} &= 
    \Gamma_0\,\frac{\alpha}{\pi}\,\frac{1}{360}\,\frac{1}{{\lambda}^2}\,
    (1 - y_0)^4\,(4 + 16\,y_0 - 5\,{y_0}^2) \\[0.2cm]
    \Big( 
    \frac{\diff \Gamma}{\diff \cos(\theta)}
    \Big)_{\rm{L}} \bigg|_{\theta = 0} &= 
    \Gamma_0\,\frac{\alpha}{\pi}\,\frac{1}{120}\,(1 - y_0)^4\,(1 + 4 y_0) \,.
\end{split} 
\end{equation}
We have found that the angular distribution 
$(\diff \Gamma / \diff \cos(\theta))_{\rm{R}}$ connected with
helicity-flip bremsstrahlung can be simulated by the following ansatz for the
differential decay spectrum (where the label ``APP'' connotes ``approximate''):
\begin{equation}
\label{APP-1}
    \Big( 
    \frac{\diff \Gamma}{\diff x\,\diff y\,\diff\!\cos(\theta)} 
    \Big)_{\rm{R}}^{\rm{APP}} =
    \Gamma_0\,\frac{2\,\alpha}{\pi}\,{\lambda}^2\,y\,(3 - 2\,x - 2\,y)\,
    \frac{1}{{( x - \cos(\theta)\,\ma) }^2} \,.
\end{equation}
When integrated over $\cos(\theta)$, this spectrum, in the limit 
$\lambda \to 0$, yields the Dalitz pair distribution given in
Eq.(\ref{limit1}). After integration over $x$ and $y$ we obtain an
angular distribution $(\diff \Gamma / \diff
\cos(\theta))_{\rm{R}}^{\rm{APP}}$ which agrees with the exact
calculation over a wide range of angles.
This formula is particularly simple for
$y_0 = 0$:
\begin{equation} 
\label{EPA-limit-1}
    \Big( 
    \frac{\diff \Gamma (y_0 = 0)}{\diff\!\cos(\theta)} 
    \Big)_{\rm{R}}^{\rm{APP}} \xrightarrow{\lambda \to 0} 
    \Gamma_0\,\frac{\alpha}{\pi}\,\frac{5}{6}\,\lambda\,
    \Big[
    1 + (\pi - \theta)\,
    \frac{\cos(\theta)}{\sin(\theta)}
    \Big]\,
    \frac{1}{\sin^2 (\theta)} \,,
\end{equation} 
which agrees with the small angle result given in Eq.(\ref{CT-lim-R-1}).
Similarly, in the forward direction ($\theta = 0$), for small values of
$\lambda$, the approximate representation Eq.(\ref{APP-1}) reproduces
the result given in Eq.(\ref{coll-lim-R-1}).

\section{Conclusions}

The purpose of this paper was to examine the dependence of the
radiative decay $\mu^- \to e^- \bar{\nu}_e \nu_\mu \gamma$ on the
helicity of the final electron, in order to demonstrate the presence
of right-handed electrons, even in the chiral limit $m_e \to 0$.
This non-intuitive feature is a consequence of helicity-flip
bremsstrahlung in QED, and was discussed in \cite{Sehgal:2003mu}
with the help of the helicity-flip splitting function derived in 
\cite{Falk:1993tf}. We have now calculated the complete
helicity-dependent spectrum of this decay from first principles,
without neglecting the electron mass, and have shown that the results of 
\cite{Sehgal:2003mu} are reproduced in the limit $\lambda \to 0$.
The presence of right-handed electrons manifests itself in the
longitudinal polarization 
$P_{\it{L}} = (\Gamma_{\rm{R}} - \Gamma_{\rm{L}}) /
(\Gamma_{\rm{R}} + \Gamma_{\rm{L}})$ which deviates from the
naive V-A value $P_{\it{L}} = -1$ even in the limit
$\lambda \to 0$. The dependence of the polarization on the photon
energy cut $y_0$, Fig.(\ref{polplot2}), shows that right-handed electrons 
occur predominantly with high energy photons.
A characteristic signature of wrong-helicity electrons is also present in the
opening-angle distribution  $\diff \Gamma / \diff \cos(\theta)$,
where the collinear region 
$\theta \lesssim \lambda / (1 - y_0)$ is dominated by right-handed
electrons. The fact that helicity-flip bremsstrahlung in the chiral
limit $m_e \to 0$ can be exactly reproduced by the simple and
universal splitting function $D_{\rm{hf}}(z)$ in many different processes,
is suggestive of a relationship to the chiral anomaly in QED
\cite{Dolgov_Zakharov_Huang_Horejsi,Freund:1994ti,Carlitz:ab}.

\def\appendix{\par
 \setcounter{section}{0}
 \setcounter{subsection}{0}
 \def\thesection{Appendix \Alph{section}}
 \def\thesubsection{\Alph{section}.\arabic{subsection}}
 \def\theequation{\Alph{section}.\arabic{equation}}
 \setcounter{equation}{0}}

\appendix

\section{}

\begin{equation}
\begin{split}
\label{PS-RL}
    \Big( \frac{\diff \Gamma}{\diff y} \Big)_{\rm{R+L}} =\,\,
    &\int_{2\,\lambda}^{x_{\rm{m}} (y)} \diff x \,
    \Big( \frac{\diff \Gamma}{\diff x\,\diff y} \Big)_{\rm{R+L}} \\ 
    =\,\, &\Gamma_0\,\frac{\alpha}{36\,\pi}\,\frac{1}{y}\,
    \Big\{
    \big[
    (-102 + 46\,y - 95\,y^2 + 109\,y^3 - 45\,y^4 + 21\,y^5 - 6\,y^6)\,(1 - y) \\
    &\phantom{\Gamma_0\,\frac{\alpha}{36\,\pi}\,\frac{1}{y}\,\Big\{\big[}
    + 6\,\lambda^2 (64 - 250\,y + 386\,y^2 - 240\,y^3 + 67\,y^4 - 20\,y^5 + 5\,y^6)\,(1 - y)^{-1}  \\
    &\phantom{\Gamma_0\,\frac{\alpha}{36\,\pi}\,\frac{1}{y}\,\Big\{\big[}
    + 18\,\lambda^4\,y\,(18 + 12\,y - 40\,y^2 + 11\,y^3 + 3\,y^4)\,(1 - y)^{-2}  \\
    &\phantom{\Gamma_0\,\frac{\alpha}{36\,\pi}\,\frac{1}{y}\,\Big\{\big[}
    + 2\,\lambda^6\,(-192 + 622\,y - 856\,y^2 + 552\,y^3 - 105\,y^4 - 21\,y^5)\,(1 - y)^{-4}  \\
    &\phantom{\Gamma_0\,\frac{\alpha}{36\,\pi}\,\frac{1}{y}\,\Big\{\big[}
    + 3\,\lambda^8\,(34 - 64\,y - 13\,y^2 + 4\,y^3)\,(1 - y)^{-4}
    \big]  \\
    &\phantom{\Gamma_0\,\frac{\alpha}{36\,\pi}\,\frac{1}{y}\,\Big\{}
    + 
    \big[
    24\,(-3 + 2\,y - 4\,y^2 + 2\,y^3)\,(1 - y) 
    + 72\,\lambda^2\,y\,(-13 + 19\,y - 7\,y^2)\,(1 - y)^{-1} \\
    &\phantom{\Gamma_0\,\frac{\alpha}{36\,\pi}\,\frac{1}{y}\,\Big\{\big[}
    + 72\,\lambda^4\,(12 - 27\,y + 18\,y^2 - 5\,y^3)\,(1 - y)^{-1} 
    + 24\,\lambda^6\,y\,(5 + 15\,y - 18\,y^2)\,(1 - y)^{-3} \\
    &\phantom{\Gamma_0\,\frac{\alpha}{36\,\pi}\,\frac{1}{y}\,\Big\{\big[}
    + 72\,\lambda^8\,(-1 + 2\,y)\,(1 - y)^{-4}
    \big]\,
    \log \Big( \frac{\lambda}{1 - y} \Big)
    \Big\} \,.
\end{split}
\end{equation}

\vspace{0.2cm}

\begin{equation}
    \Big(
    \frac{\diff \Gamma}{\diff x} 
    \Big)_{\rm{R+L}} = 
    \Gamma_0\,\frac{\alpha}{\pi}\,\frac{1}{\ma}\,u^{-4}\,(u^2 - {\lambda}^2)
    \bigg[
    k_1 + \log \Big( \frac{u}{\lambda} \Big)\,k_2 +
    \log \Big( \frac{1 - u}{y_0} \Big)\,k_3 +
    \log \Big( \frac{u}{\lambda} \Big)\,\log \Big( \frac{1 - u}{y_0} \Big)\,k_4
    \bigg] \,,
\end{equation}
where $u = (x + \ma)/2$, and
\begin{equation}
\begin{split}
\label{xi2-u-1}
    k_1 \,:=\,\,
    &\frac{1}{36}\,(u^2 - {\lambda}^2)\,(1 - y_0 - u)
    \Big[
    u^2\,
    \big(
    -300 + 399\,u + 71\,u^2 + 2\,u^3 + 8\,u^4 + 132\,y_0 + 63\,u\,y_0 - 10\,u^2\,y_0 \\
    &\phantom{\frac{1}{36}\,(u^2 - {\lambda}^2)\,(1 - y_0 - u)\Big[u^2\,\big(}
    - 8\,u^3\,y_0 + 24\,{y_0}^2 - 12\,u\,{y_0}^2 + 8\,u^2\,{y_0}^2
    \big) \\
    &\phantom{\frac{1}{36}\,(u^2 - {\lambda}^2)\,(1 - y_0 - u)\Big[}
    + {\lambda}^2\,u\,
    \big(
    555 - 259\,u - 61\,u^2 - 19\,u^3 + 87\,y_0 - 28\,u\,y_0 + 19\,u^2\,y_0 \\
    &\phantom{\frac{1}{36}\,(u^2 - {\lambda}^2)\,(1 - y_0 - u)\Big[+ {\lambda}^2\,u\,\big(}
    - 12\,{y_0}^2 + 8\,u\,{y_0}^2
    \big) \\
    &\phantom{\frac{1}{36}\,(u^2 - {\lambda}^2)\,(1 - y_0 - u)\Big[}
    + {\lambda}^4\,
    \big( 
    122 - 67\,u - 19\,u^2 - 22\,y_0 + 19\,u\,y_0 + 8\,{y_0}^2
    \big) 
    \Big] \,, \\
    k_2 \,:=\,\,
    &\frac{1}{3}\,u\,(1 - y_0 - u)\,
    \Big[
    u^2\,
    \big( 
    5 + 17\,u - 34\,u^2 + 5\,y_0 - 14\,u\,y_0 - 4\,{y_0}^2
    \big)
    + 3\,{\lambda}^2\,u\,
    \big( 
    6 - 19\,u + u^2 - 6\,y_0 - 5\,u\,y_0
    \big) \\
    &\phantom{\frac{1}{3}\,u\,(1 - y_0 - u)\,\Big[}+ 12\,{\lambda}^4\,	
    \big( 
    -4 - u + 3\,u^2
    \big)		
    \Big] \,, \\
    k_3 \,:=\,\,
    &4\,({\lambda}^2 - u^2)\,
    \big[
    u^3\,(3 - 2\,u) + {\lambda}^2\,u\,(3 - 8\,u + 3\,u^2) + {\lambda}^4\,(-2 + 3\,u)
    \big] \,, \\
    k_4 \,:=\,\,
    &4\,({\lambda}^2 + u^2)\,
    \big[
    u^3\,(3 - 2\,u) + {\lambda}^2\,u\,(3 - 8\,u + 3\,u^2) + {\lambda}^4\,(-2 + 3\,u)
    \big] \,,
\end{split}
\end{equation}

\vspace{0.2cm}

\begin{equation} 
\begin{split}
    \Big( \frac{\diff \Gamma}{\diff y} \Big)_{\rm{R}}
    \,=\, &\int_{2\,\lambda}^{x_{\rm{m}}(y)} \diff x\, 
    \Big( \frac{\diff \Gamma}{\diff x\,\diff y} \Big)_{\rm{R}} \\
    \,=\, &\Gamma_0\,\frac{\alpha}{\pi}\, 
    \Big[
    y\,(1 - y)\,(2 - y) + h_1 + 
    \log \Big( \frac{1 - y}{\lambda} \Big)\,h_2 +
    \log^2 \Big( \frac{1 - y}{\lambda} \Big)\,h_3 
    \Big] \,, \\
\end{split} 
\end{equation}
where
\begin{equation}
\begin{split}
\label{photon-spectrum-R-eq}
    h_1 \,:= \,\,
    &4\,\lambda\,y\,(-3 + 2\,y) + \frac{1}{18}\,{\lambda}^2\,\frac{1}{y}\,\frac{1}{1 - y}\,
    (-248 + 176\,y + 489\,y^2 - 304\,y^3 - 53\,y^4 - 6\,y^5) \\
    &+ \frac{8}{9}\,{\lambda}^3\,\frac{1}{y}\,
    (48 - 10\,y + 15\,y^2 + 3\,y^3) + \frac{1}{4}\,{\lambda}^4\,\frac{1}{y}\,\frac{1}{(1 - y)^2}\,
    (-150 + 320\,y - 209\,y^2 + 16\,y^3 + 29\,y^4 - 2\,y^5) \\  
    &+ \frac{4}{9}\,{\lambda}^5\,\frac{1}{y}\,
    (32 - 12\,y - 3\,y^2) + \frac{1}{18}\,{\lambda}^6\,\frac{1}{y}\,\frac{1}{(1 - y)^3}\, 
    (-120 + 412\,y - 465\,y^2 + 114\,y^3 + 18\,y^4) \\
    &+ \frac{1}{36}\,{\lambda}^8\,\frac{1}{y}\,\frac{1}{(1-y)^4}\,
    (38 - 140\,y - 21\,y^2 + 6\,y^3) \,, \\
    h_2 \,:=\,\,
    &\frac{2}{3}\,{\lambda}^2\,\frac{1}{y}\,\frac{1}{1 - y}\,
    (8 + 13\,y - 30\,y^2 + 7\,y^3 + 5\,y^4)
    + {\lambda}^4\,\frac{1}{y}\,\frac{1}{1 - y}\,(-16 + 22\,y - 19\,y^2 + 3\,y^3) \\
    &+ \frac{4}{3}\,{\lambda}^6\,\frac{1}{(1 - y)^3}\,(-1 - 6\,y + 6\,y^2) 
    + \frac{2}{3}\,{\lambda}^8\,\frac{1}{y}\,\frac{1}{(1 - y)^4}\,(1 - 4\,y) \,, \\
    h_3 \,:=\,\, 
    &2\,{\lambda}^2\,(1 - y)\,(-3 - y) +
    2\,{\lambda}^4\,\frac{1}{y}\,(-2 - y + y^2) \,.
\end{split} 
\end{equation}

\vspace{0.2cm}

\begin{equation} 
    \Big(
    \frac{\diff \Gamma}{\diff x} 
    \Big)_{\rm{R}} = 
    \Gamma_0\,\frac{\alpha}{\pi}\,
    \bigg[
    j_1 + \log \Big( \frac{u}{\lambda} \Big)\,j_2 +
    \log \Big( \frac{1 - u}{y_0} \Big)\,j_3 +
    \log \Big( \frac{u}{\lambda} \Big)\,\log \Big( \frac{1 - u}{y_0} \Big)\,j_4
    \bigg] \,, 
\end{equation} 
where
\begin{equation} 
\begin{split}
\label{PSR-1}
    j_1 \,:=\,\,
    &\frac{1}{36}\,\frac{1}{u^3}\,(1 - y_0 - u)\,
    \Big\{
    u^3\,
    \big[ 
    30 + 12\,u^2 + u\,(-42 - 12\,y_0)  + 30\,y_0 - 24\,{y_0}^2 
    \big] \\
    &\phantom{\frac{1}{36}\,\frac{1}{u^3}\,(1 - y_0 - u)\,\Big\{}
    + \lambda^2\,u^2\, 
    \big[
    264 - 168\,y_0 - 24\,{y_0}^2 +
    u\,(261 - 15\,y_0 - 12\,{y_0}^2) \\
    &\phantom{\frac{1}{36}\,\frac{1}{u^3}\,(1 - y_0 - u)\,\Big\{ + \lambda^2\,u^2\,}
    + u^2\,(-155 - 26\,y_0 + 4\,{y_0}^2) + u^3\,(-5 + 5\,y_0) + u^4\,(-5) 
    \big] \\
    &\phantom{\frac{1}{36}\,\frac{1}{u^3}\,(1 - y_0 - u)\,\Big\{}
    + \lambda^4\,u\,
    \big[
    -555  - 87\,y_0 + 12\,{y_0}^2 + u\,(199 - 32\,y_0 + 4\,{y_0}^2) \\
    &\phantom{\frac{1}{36}\,\frac{1}{u^3}\,(1 - y_0 - u)\,\Big\{ + \lambda^4\,u\,}
    + u^2\,(73 + 5\,y_0) + u^3\,(-5)
    \big] \\
    &\phantom{\frac{1}{36}\,\frac{1}{u^3}\,(1 - y_0 - u)\,\Big\{}
    + \lambda^6\,
    \big[
    -122 + 22\,y_0 - 8\,{y_0}^2 + 10\,u\,(4 - y_0) + 10\,u^2
    \big]
    \Big\} \,, \\ 
    j_2 \,:=\,\, &\frac{1}{3}\,\frac{1}{u^2}\,\frac{{\lambda}^2}{u^2 - {\lambda}^2}\,
    \Big\{
    u^2\,(1 - y_0 - u)\,
    \big[
    -5 - 5\,y_0 + 4\,{y_0}^2 + u\,(-19 + 12\,y_0 + 4\,{y_0}^2) + u^2\,(-13 - y_0) + 13\,u^3
    \big] \\
    &\phantom{\frac{1}{3}\,\frac{1}{u^2}\,\frac{{\lambda}^2}{u^2 - {\lambda}^2}\,\Big\{}
    + 6\,{\lambda}^2\,u\,(1 - y_0 - u)\,
    \big[
    -3 + 3\,y_0 + u\,(2 + 3\,y_0) + u^2\,(-4 + y_0) + u^3
    \big] \\
    &\phantom{\frac{1}{3}\,\frac{1}{u^2}\,\frac{{\lambda}^2}{u^2 - {\lambda}^2}\,\Big\{}
    + 24\,{\lambda}^4\,(2 - u)(1 + u)\,(1 - y_0 - u)
    \Big\} \,, \\
    j_3 \,:=\,\, &4\,\frac{1}{u^3}\,{\lambda}^2\,(1 - u)\,
    ({\lambda}^2 - u^2)\,(-2\,{\lambda}^2 + 3\,u - u^2) \,, \\
    j_4 \,:=\,\, &4\,\frac{1}{u^3}\,{\lambda}^2\,(1 - u)\,
    ({\lambda}^2 + u^2)\,(-2\,{\lambda}^2 + 3\,u - u^2) \,.
\end{split} 
\end{equation}

\end{document}